\documentclass{aa}
\usepackage{graphicx}
\begin{document}
%
\def\hi {H\,{\sc i}}
\def\oiii {O{\sc iii}}
\def\hdueo {H$_2$O}
\def\fri {FR\,I}
\def\frii {FR\,II}
\def\txs {TXS\,2226{\tt -}184}
\def\pks {PKS\,2322{\tt -}123}
\def\radm {rad m$^{-2}$}
\def\ab {$\sim$}
\def\etal {{\sl et~al.\ }}
\def\dg{$^{\circ}$}
\def\kms{km\,s$^{-1}$}
\def\solmass {\hbox{M$_{\odot}$}}
\def\solum {\hbox{L$_{\odot}$}} 
\def\ffam {\hbox{$\,.\!\!^{\prime}$}}
\def\ffas {\hbox{$\,.\!\!^{\prime\prime}$}}
\def\ffM {\hbox{$\,.\!\!^{\rm M}$}}
\def\ffm {\hbox{$\,.\!\!^{\rm m}$}}
\def\ffs {\hbox{$\,.\!\!^{\rm s}$}}
\title{The innermost region of the water megamaser radio galaxy 3C~403}


\author{A.\ Tarchi \inst{1,2}
  \and
   A.\ Brunthaler \inst{3}
  \and
  C.\ Henkel \inst{3}
  \and
  K.\ M.\ Menten \inst{3}
  \and
  J.\ Braatz \inst{4}
  \and
  A.\ Wei{\ss} \inst{3,5}}

\offprints{A. Tarchi}

\institute{INAF-Osservatorio Astronomico di Cagliari, Loc. Poggio dei Pini, Strada 54, 09012 Capoterra (CA), Italy\\
 \email{atarchi@ca.astro.it}
 \and
 INAF-Istituto di Radioastronomia, Via Gobetti 101, 40129 Bologna, Italy
 \and
 Max-Planck-Institut f{\"u}r Radioastronomie, Auf dem H{\"u}gel 69, D-53121 Bonn, Germany
 \and
 National Radio Astronomy Observatory, 520 Edgemont Road, Charlottesville, VA 22903
 \and
 IRAM, Avenida Divina Pastora 7, 18012 Granada, Spain}

\date{Received ; accepted}

\abstract
{The standard unified scheme of active galactic nuclei requires the presence of high column densities of gas and dust potentially obscuring the central engine. So far, few direct subarcsecond resolution studies of this material have been performed toward radio galaxies.}
{The goal of this paper is to elucidate the nuclear environment of the prototypical X-shaped Fanaroff-Riley type II radio galaxy 3C\,403, the only powerful radio galaxy known to host an H$_2$O megamaser.} 
{Very Large Array A-array and single-dish Green Bank and Effelsberg 1.3\,cm measurements were performed to locate and monitor the water maser emission. Very Long Baseline Interferometry 6\,cm continuum observations were taken to analyze the spatial structure of the nuclear environment at even smaller scales, while the CO $J$=1--0 and 2--1 transitions were observed with the IRAM 30-m telescope to search for thermal emission from a spatially extended, moderately dense gas component.}
{Positions of the H$_2$O maser features and the continuum emission from the core coincide within 5 mas (5.5\,pc). Intensities of the two main maser components with (isotropic) luminosities sometimes surpassing 1000\,L${_\odot}$ appear to be anti-correlated, with typical timescales for strong variations of one year. If the variations are intrinsic to the cloud(s), the implied angular source size would be $\la$0.3\,mas and the brightness temperature $\ga$5$\times$10${^8}$\,K. The VLBI continuum observations support a scenario where a nuclear core, represented by the dominant central radio continuum component, is accompanied by a jet and counterjet, directed toward the western and eastern large scale lobes of the galaxy. CO remains undetected, providing a maximum scale size of $\sim$50\,pc $\times$ (500\,K/$T_{\rm b})^{1/2}$, with $T_{\rm b}$ denoting the brightness temperature of the CO $J$=1--0 line. Possible scenarios that could produce the observed maser emission are outlined. Adopting a mass of several 10$^{8}$ for the nuclear engine, the observed maser features can only be interpreted in terms of an accretion disk as in NGC\,4258, if they solely represent the systemic velocity components. The receding and approaching parts of the putative maser disk are, however, not seen and a secular velocity drift of the observed features is not (yet) apparent. Most likely, the two main maser components mark shocked molecular gas interacting with the nuclear jets. The X-shaped morphology of the radio galaxy may point at a binary nuclear engine. This possibility, greatly complicating the nuclear environment of 3C\,403, should motivate a number of worthwhile follow-up studies.}
{}

\keywords{Galaxies: individual: 3C\,403 -- 
     Galaxies: active -- 
     Galaxies: ISM -- 
     masers -- 
     Radio lines: ISM -- 
     Radio lines: galaxies}

\titlerunning{The innermost region of 3C~403}
\authorrunning{Tarchi et al.}

\maketitle


\section{Introduction}

The physical conditions in active galactic nuclei (AGN) are unique in the cosmos. Stellar and gas densities are exceptionally 
large and enormous amounts of energy and angular momentum are released as the material accretes onto the supermassive nuclear
engine. Performing studies of the structure, kinematics and excitation of this material is the sole mean available to investigate (super)massive ultracompact objects. 

Because of obscuration and crowding, only a few molecular tracers can provide direct information on the innermost regions of active galaxies. 
The H$_2$O\ $\lambda$=1.3\,cm maser line, profiting from a long wavelength (no obscuration by dust grains), from its enormous 
luminosity (up to 10$^{4}$\,L$_{\odot}$  (isotropically), corresponding to $\sim$10$^{53}$ photons/s; see Koekemoer et al. 
\cite{koekemoer95}; Barvainis \& Antonucci \cite{barvainis05}) and the availability of Very Long Baseline Interferometry (VLBI) 
networks, is a particularly suitable tracer of this environment. The line samples gas with kinetic temperatures of several 100\,K 
and densities of $\sim$10$^{8}$\,cm$^{-3}$ (e.g., Kylafis \& Norman \cite{kylafis87}, \cite{kylafis91}). 

So-called "disk-masers" as in NGC\,4258 allow us to map nuclear accretion disks (e.g. Greenhill et al. \cite{greenhill95}; Miyoshi 
et al. \cite{miyoshi95}; Herrnstein et al. \cite{herrnstein05}). Analysis of the dynamics of observed Keplerian disks leads to estimates of the mass of the nuclear engine and to a determination of the distance to the galaxy, independent 
of any standard candles. So-called "jet-masers" as in Mrk~348 (Peck et al. \cite{peck03}) are associated with the pc-scale jets
and  provide estimates, through reverberation mapping, of the speed of the material in the jet. A further class of masers is 
associated with nuclear outflows (Circinus; Greenhill et  al. \cite{greenhill03}). 

Measuring the proper motion of galaxies has 
been another long-standing problem. The solution is to find sufficiently luminous ultracompact sources that can be observed with 
VLBI techniques involving phase referencing. Extragalactic H$_2$O masers are suitable targets and could provide, through their 
measured radial velocities and proper motions, three dimensional velocity vectors in space (Brunthaler et al. \cite{brunthal05}, 
\cite{brunthal07}), not only for the masers themselves but also for the nuclei of their parent galaxies. 

Based on the assumption that the water masers are located in circumnuclear tori, (often) aligned with the nucleus and amplifying its continuum, linearly for saturated masers and exponentially for unsaturated ones, one should expect a high detection rate in radio-loud galaxies (in particular those with the radio axis close to the plane of the sky). However, water megamasers have been found so far only in low-radio power AGN ($\rm P_{1GHz} < 10^{22-23}~W\,Hz^{-1}$), mostly in galaxies classified as Sy~2s or LINERs (Braatz et al. \cite{braatz96}, \cite{braatz97}).  



A first systematic search for the $\lambda$=1.3\,cm line in radio galaxies was performed by Henkel et al. (\cite{henkel98}). 
No maser was detected in a sample of $\sim$50 Fanaroff-Riley type I (FR~I) galaxies. Subsequently, targets belonging to the more powerful Fanaroff-Riley type II (FR~II) class 
were observed by two groups (Tarchi et al., Lara et al., unpublished) again with negative results.

According to the current paradigm for radio-loud AGN, FR~II galaxies host type 2 AGN with narrow optical 
emission lines and prominent, extended nuclear jets. Since the jets are ejected at a small angle with respect to the plane 
of the sky, FR~II galaxies should contain a nuclear torus being viewed approximately edge-on (for a review, see e.g. Urry \& 
Padovani \cite{urry95}). So far, however, the presence and nature of these hypothesized tori has remained an enigma. The nuclear 
region of the prototypical FR~II galaxy Cyg~A shows evidence for dust and obscuration with gas column densities up to several 
10$^{23}$\,cm$^{-2}$.  Morphology and physical parameters of the nuclear region as well as the mechanism triggering nuclear 
activity are, however, not yet fully understood (e.g., Conway \& Blanco \cite{conway95}, Carilli \& Barthel \cite{carrilli96}; 
Fuente et al.\ \cite{fuente00}; Bellamy \& Tadhunter \cite{bellamy04}).

Very recent modifications to the standard unified scheme (hereafter SUS) state that FR~I galaxies (or at least the majority of 
them) seem to lack a geometrically and optically thick molecular torus. This scenario is now supported by an increasing number 
of studies (e.g., Chiaberge et al. \cite{chiaberge99}; Perlman et al. \cite{perlman01}; Verdoes Kleijn et al. \cite{verdoes02}; 
Whysong \& Antonucci \cite{whysong04}). In order to account for the absence of broad emission lines in narrow--lined FR~IIs 
(within the framework of the SUS believed to represent the parent population of radio-loud quasars) a geometrically and 
optically thick obscuring layer is required instead (Barthel \cite{barthel89} and references therein).

Recently, a luminous ($L_{\rm H_2O}$$\sim$10$^3$\,L$_{\odot}$) H$_2$O megamaser has been detected in 3C\,403 (Tarchi et al. 
\cite{tarchi03}; hereafter THC), an FR~II galaxy at roughly the same distance as Cyg~A. 3C\,403 is the first and so far only 
 powerful radio galaxy with a detected H$_2$O maser.

The kpc radio morphology of 3C\,403 is typical for X-shaped radio galaxies. The cause of the ``X'' radio shape is still debated. Arguments exist that support the origin of this peculiar morphology as due to a sudden change in the jet direction. Merritt \& Eckers (\cite{merritt02}) have argued that the sudden change in the jet direction reflects a sudden ($<$ few yr) change in the black hole spin axis due to a black hole merger (see also Sect.\,5.2.2). Alternative models imply instead slow precession of the jet axis or lobe backflow (see Dennett-Thorpe et al. \cite{dennett02}, Schoenmakers et al. \cite{schoen00}). 

The Very Large Array (VLA) $\lambda$=3.6\,cm maps of 
Black et al. (\cite{black92}; hereafter BBL; their Fig.\,13) and Dennett-Thorpe et al. (\cite{dennett02}; hereafter DSL; their 
Fig.\,1) show two bright radio lobes towards the east and west with hot-spots and two weaker, longer and broader wings to the 
north-west and south-east. The two bright lobes are caused by jets containing a number of prominent knots. Adopting a WMAP cosmology 
($H_0$=71, $\Omega_{\rm M}$=0.27, $\Omega_\lambda$=0.73; Spergel et al.\ \cite{spergel03}), the redshift of $z$=0.059 leads to
an angular size distance of $D_{\rm A}$ $\sim$ 230\,Mpc, at which 1\,mas corresponds to $\sim$1.1\,pc.

In this paper, we present results from a follow-up VLA A-array observation of the maser emission and the outcome of a search for 
CO emission using the 30-m IRAM telescope at Pico Veleta. We also discuss monitoring observations of the maser line spanning two years. Furthermore, to obtain a more realistic picture of the innermost regions of 3C\,403 and to find out 
whether an accretion disk scenario is  plausible, we mapped the 6 cm radio continuum emission with parsec scale resolution using 
the European VLBI Network (EVN).

\section{Observations and data reduction}\label{obssect}

\noindent{\bf Effelsberg} 3C~403 was observed in the $6_{16} - 5_{23}$ transition of H$_2$O (rest frequency: 22.23508\,GHz) with the 
100-m telescope of the MPIfR at Effelsberg\footnote{The 100-m telescope at Effelsberg is operated by the Max-Planck-Institut f{\"u}r 
Radioastronomie (MPIfR) on behalf of the Max-Planck-Gesellschaft (MPG).} at four epochs between January 2003 and April 2004. The beam 
width was 40$''$. The observations were made with a 18--26\,GHz HEMT receiver that was sensitive to the two orthogonal linear polarizations.
A dual beam switching mode with a beam throw of 2\arcmin\ and a switching frequency of $\sim$1\,Hz was chosen. The Effelsberg AK90 
spectrometer was used as backend. System temperatures, including atmospheric contributions, were $\sim$40--60\,K on an antenna temperature 
scale ($T_{\rm A}^*$). The beam efficiency was $\eta_{\rm b}$$\sim$0.5. Flux calibration was obtained by measuring W3(OH) (see Mauersberger 
et al. \cite{mauer88}). Gain variations of the telescope as a function of elevation were taken into account (see e.g., Eq.\,1 of Gallimore 
et al. (\cite{gallimore01}). The pointing accuracy was better than 10\arcsec. All data were reduced using standard procedures of the 
GILDAS software package (http://www.iram.fr/IRAMFR/GS/gildas.htm). 

\hfill\break\noindent
{\bf GBT} Observations were made with the Green Bank Telescope (GBT) of the NRAO\footnote{The National Radio Astronomy Observatory 
(NRAO) is a facility of the National Science Foundation operated under cooperative agreement by Associated Universities, Inc.} during six 
sessions between 2003 October 20, and 2005 February 15. We used the 18--22\,GHz K-band receiver, which has two beams at a fixed separation 
of 3$^{\prime}$ in azimuth. The GBT beamwidth is 35\arcsec\ at the red shifted maser frequency of 21\,GHz, and pointing uncertainties were better than 10\arcsec. The 
data were taken in total power mode, and the telescope was nodded between two positions on the sky such that the source was always in one 
of the two beams during integration. We used a nod cycle of 2 minutes per position.

\begin{figure}
\centering
\includegraphics[width= 8 cm]{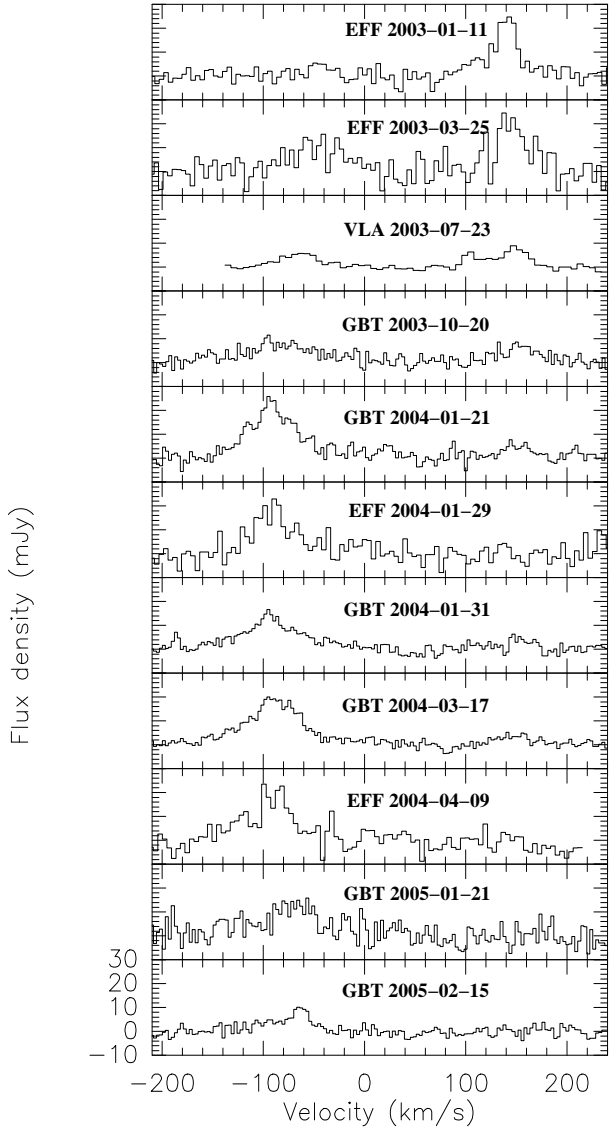}
\caption{Maser lines in 3C\,403 observed with Effelsberg (four epochs), the GBT (6 epochs), and the VLA (one epoch). The spectrum
from the first epoch (upper panel) was already published in THC. The channel spacing is 312\,kHz ($\cor$ 4.5\,km\,s$^{-1}$), 195\,kHz 
($\cor$ 2.8\,km\,s$^{-1}$), and 391\,kHz ($\cor$ 5.6\,km\,s$^{-1}$) for the spectra taken with Effelsberg, the GBT, and the VLA, 
respectively. Zero velocity corresponds to a frequency of 20996.28 MHz (c$z$ = 17688\,km\,s$^{-1}$ w.r.t. the Local Standard of Rest 
(LSR)), the galaxy's recessional velocity according to the NASA/IPAC database (NED)}
\label{monitor}
\end{figure}

The spectrometer was configured with two 200\,MHz bandpasses, one centered on the systemic velocity of 3C\,403 and the second redshifted 
by 180\,MHz. No emission was detected beyond what is shown in Fig.\,\ref{monitor}. The zenith system temperature was between 35 and 45\,K. 
Atmospheric opacity was estimated using system temperature and weather data, and ranged from 0.04 to 0.06 at the zenith. We reduced the data using GBTIDL.

\hfill\break\noindent
{\bf VLA} Water maser emission was observed at the red shifted maser frequency of 21\,GHz on July 23 and 25, 2003, with the Very Large Array (VLA) of the NRAO$^{2}$ in 
its A configuration for an 8 hour total on-source time. We observed with two 12.5\,MHz IFs centered on the two maser features detected 
by Tarchi et al. (\cite{tarchi03}; see the upper spectrum of Fig.\,\ref{monitor}). Each IF was split into 32 channels providing a 
channel spacing of $\sim$ 6 \kms. The source 1331+305 (2.66\,Jy) was used as flux and bandpass calibrator. The point source 1950+081, 
at a distance of 5\fdg6 from 3C\,403 (there was no suitable closer source), was used for phase calibration. The two IFs were calibrated 
separately and then combined (AIPS task: UVGLU) yielding a total velocity coverage of $\sim$360\,km\,s$^{-1}$. The data were 
Fourier-transformed using natural weighting to create a $256\times256\times64$ data cube. The radio continuum was subtracted using the 
AIPS task UVLSF. This task fits a straight line to the real and imaginary parts of selected channels and subtracts the fitted baseline 
from the spectrum, optionally flagging data having excess residuals. In addition, it provides the continuum as a UV data set, which 
has been used to create continuum maps of the galaxy. The restoring beam is $0\,\farcs1\times0\,\farcs1$. The rms noise per channel 
becomes 0.6 mJy, consistent with the expected thermal noise.

\hfill\break\noindent
{\bf EVN} 3C\,403 was observed on May 20, 2004, with the European VLBI Network\footnote{The European VLBI Network is a joint facility of European, Chinese, South African and other radio astronomy institutes funded by their national research councils.} (EVN) at 6\,cm. Eight 8\,MHz bands, each at 
right and left circular polarization, were employed. The total observing time was 10.5 hours. We used 4C\,+02.49 as a phase-reference 
source and switched every 2 minutes between the two sources. The initial calibration was performed with the AIPS package. A priori  
amplitude calibration was applied using system temperature measurements and standard gain curves. Fringes were found in the 3C\,403 
data itself on all baselines, so we used this source as phase-reference and applied the solutions also to 4C\,+02.49. The data were 
self-calibrated and mapped using the software package DIFMAP (Shepherd et al.\ \cite{shepherd94}). We started with phase-only 
self-calibration and later included phase-amplitude self-calibration with solution intervals slowly decreasing down to one minute. 

\hfill\break\noindent
{\bf Pico Veleta} 3C\,403 was observed in a search for CO emission on November 6, 2004, using the IRAM\footnote{IRAM is supported by INSU/CNRS (France), the MPG (Germany), and the IGN (Spain).} 30-m telescope at Pico Veleta, 
Spain. Using the A/B receiver combination we searched simultaneously for CO(1--0) and CO(2--1) with the A/B\,100 receivers tuned 
to 108.85\,GHz and the A/B\,230 receivers tuned to 217.69\,GHz. Beamwidths were 23\arcsec\ and 13\arcsec, respectively. Spectra were 
obtained using the wobbler switch technique with a beam throw of 60\arcsec\ and a switching frequency of 0.5\,Hz. The data were 
recorded using the 1\,MHz (512 channels, 1\,MHz channel spacing) and 4\,MHz (256 channels, 4\,MHz channel spacing) filterbanks
for the 109 and 218\,GHz observations, respectively. The effective velocity coverage is 1400 km/s at both frequencies. Typical 
system temperatures were 160 and 500\,K ($T_{\rm A}^*$). The fast switching procedure and the good quality of the resulting 
baselines also provide a rough measure of continuum levels (see Sect.\,4.1.3). 3C\,403 was observed for about 70 minutes (on+off) 
leading to rms noise levels of 4.2 and 12.5\,mJy after smoothing the 109 and 218\,GHz spectral bands to channel widths of 44 and 
50\,km\,s$^{-1}$.

\begin{figure}
\centering
\includegraphics[width= 8 cm]{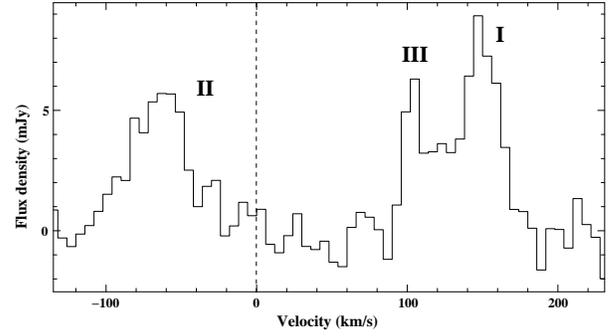}
\caption{The main water maser features in 3C\,403 observed with the VLA A-array on July 23, 2003 (see Sect.\,\ref{vlares} and 
Fig.\,\ref{monitor}). The dashed vertical line marks the nominal systemic velocity of the galaxy, V$_{\rm sys}$ = c$z$ = 
17688\,km\,s$^{-1}$.}
\label{spectrum}
\end{figure}

\begin{figure}
\centering
\includegraphics[width= 8 cm]{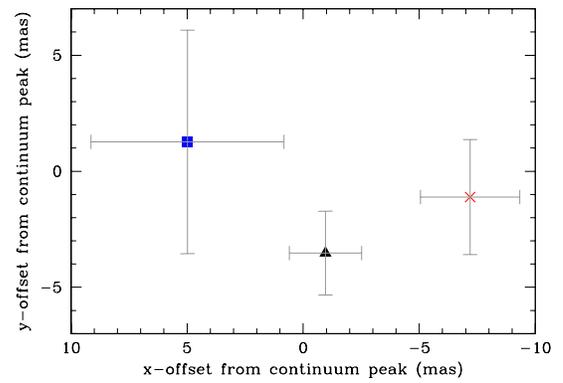}
\caption{Close-up of the region around the nucleus of 3C\,403. Shown as a filled triangle, a cross, and a filled square are the variance-weighted average locations with error bars of maser components I, II, and III, respectively. Position offsets in (RA$_{\rm J2000}$, Dec$_{\rm J2000}$) are relative to the 22-GHz continuum peak position.}
\label{gradient}
\end{figure}

\section{The systemic velocity of 3C\,403}\label{vref}

For the following sections it is of fundamental relevance to assess the accuracy of the reported systemic velocity $V_{\rm sys}$ of 3C\,403. Optical emission line measurements indicate $V_{\rm sys}$ = 17688\,km\,s$^{-1}$ ($z$ = 0.059) with no uncertainty value provided (Spinrad et al. 
\cite{spinrad85} and references therein). Previous studies indicate that recessional velocities of galaxies derived from optical emission lines tend to have a systematic uncertainty, often being smaller than systemic velocities derived from H{\sc i} (e.g. Mirabel \& Wilson \cite{mirabel84}; Morganti et al. \cite{morganti01}). This difference is attributed to a net outflow of gas in the optically unobscured front side of narrow emission line regions (e.g. Mirabel \& Wilson \cite{mirabel84}). Unfortunately, no neutral hydrogen has been so far detected in 3C~403 (see Sect.~5.1.2) to perform a comparison with the optically-derived recessional velocity. Baum et al. (\cite{baum90}), while not providing a recessional velocity, derived on the basis of [O{\sc i}], [N{\sc ii}], H$\alpha$ and [S{\sc ii}] data a rotation curve of 3C\,403 that indicates no strong anomalies. 
From this we conclude that the uncertainty in $V_{\rm sys}$, as derived by Spinrad et al. \cite{spinrad85}, should not exceed 100\,km\,s$^{-1}$.

\section{Results}

\subsection{Spectral line emission} 

\subsubsection{Maser monitoring}

Figure\,\ref{monitor} shows the water maser spectra from 3C\,403 at eleven epochs from January 2003 to February 
2005, observed with the Effelsberg telescope, the GBT, and the VLA A-array. Past Effelsberg unpublished data from December 1997 showed no sign of emission at a 3$\sigma$ level of 60 mJy. Fig.\,\ref{spectrum} displays our most sensitive spectrum, that obtained with the VLA indicating the possible presence of three features that, for the sake of convenience, we indicate as component I, II, and III.
 
The most obvious change that can be deduced from Fig.~\ref{monitor} is that, in the early spectra, emission is mostly present on the red-shifted side (component I), while later most of the emission (component II) is blue-shifted w.r.t. the systemic velocity.   
In January and March 2003, component I had a peak line strength of 20--25 mJy, becoming weaker at subsequent epochs and remaining undetected in the most recent spectra ($<$5\,mJy). Component II correspondingly increases its flux from $\sim$4 to 20\,mJy before significantly fading in the last two spectra. 
From March 2003 to January 2004, component II seems to experience a first shift from about --45 to --95\,km\,s$^{-1}$ and a second one back to about \hbox{--60}\,km\,s$^{-1}$ in February 2005. Of course, the possibility that there is actually no shift and we are, instead, witnessing the brightening and fading of different components is also a possibility. Component III, detected at a low significant level only in the first three spectra, fades below the 5\,mJy noise level together with component I. 

While no systematic regular velocity drift involving all the feature can be confidently observed, it is remarkable that within about 15 months the 
maser lines swapped their intensities. A time scale of one year and peak flux densities of order 20\,mJy indicate an angular 
source size of $\la$0.3\,mas and a peak brightness temperature of $\ga$5$\times$10$^8$\,K, if variations are intrinsic to the 
cloud(s) and are not triggered by fluctuations of the continuum background. Each individual component reaches peak isotropic 
luminosities well in excess of 1000\,L$_{\odot}$. So far, only four other maser sources were reported to have a similar power, 
TXS22226--184 (Koekemoer et al. \cite{koekemoer95}), Mrk~034 (Henkel et al. \cite{henkel05}), J0804+3607 (Barvainis \& Antonucci 
\cite{barvainis05}), and UGC~5101 (Zhang et al. \cite{zhang06}). A variation by more than a factor of two has not yet been reported in such a luminous maser component. Less luminous masers with linewidths in excess of 10\,km\,s$^{-1}$ are commonly more stable.

\subsubsection{The VLA view of the H$_2$O maser}\label{vlares}

In all VLA velocity channels (see Fig.\,\ref{spectrum}; for a preliminary report, see also Tarchi et al. \cite{tarchi05}) the 
emission peaks at the position RA$_{\rm J2000}$ = 19$^{\rm h}$ 52$^{\rm m}$ 15\fs80, Dec$_{\rm J2000}$ = 
02$^{\rm \circ}$ 30$^{\rm \prime}$ 24\farcs2. The nominally estimated error for absolute positions for a VLA A-array 
map is 0\,\farcs1 when a strong nearby calibrator can be used for phase 
referencing (as done in our observations, using the source 1950+081). Since we have also used the "fast switching" technique, this error can be taken as a safe upper limit. 

The total integrated intensity (moment-0; in mJy\,km\,s$^{-1}$\,beam$^{-1}$) VLA plots of components I--III show that they are 
unresolved at the 0\,\farcs1 resolution and coincident in position with the radio continuum nucleus of the galaxy (see also 
Sect.\,\ref{continuosect}). The alignment between line and continuum peak positions derived from the same data is only limited by 
the signal-to-noise ratios (SNRs) and is estimated to be $\sigma_{\rm rel} = \sqrt{(\theta_{l}/2 \cdot SNR_{l})^{2} + 
(\theta_{c}/2 \cdot SNR_{c})^{2}}$ $=$ $\sqrt{(0.1/2 \cdot 15)^{2} + (0.1/2 \cdot 160)^{2}}$ $\sim$ 0\farcs005 $\cor$ 5.5 pc, where $\theta$ denotes the restored 
beam size, and l and c refer to line and continuum emission, respectively. 

Fig.\,\ref{gradient} presents a close-up of the region around the nucleus of 3C\,403, showing the relative distribution of maser components I, II, and III. The symbols and error bars mark the variance-weighted average locations and uncertainties of the three maser components derived fitting, with the AIPS task JMFIT\footnote{For a discussion on the accuracy of the parameters derived from JMFIT, see e.g. Henkel et al. (\cite{henkel04}; their Sect.\,3)}, the maser emission in each channel with flux density $>$\,5 times the rms noise of 0.6 mJy\,beam$^{-1}$. Our result indicates an upper limit for the offset between different components of 12 $\pm$ 4.5 mas. The red-shifted emission (components I and II) is seemingly located eastwards w.r.t. the blue-shifted one (component II).

In addition, we have produced a $\sim$ $3^{\prime}\times3^{\prime}$ map, covering the entire radio extent of the galaxy and no 
additional maser spots were detected above the 0.6\,mJy level given in Sect.\,\ref{obssect}.

\begin{figure}
\centering
\includegraphics[width=4.6cm,angle=-90]{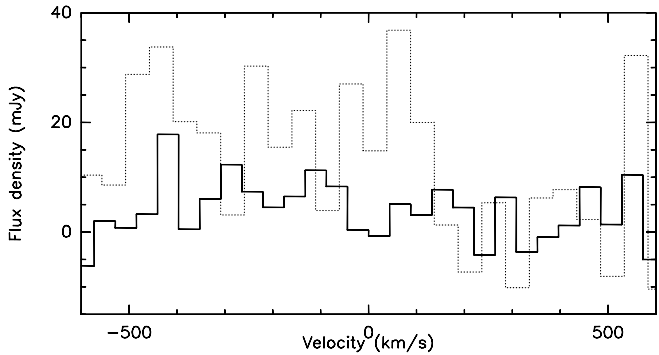}
\caption{30-m IRAM telescope CO(1--0) (solid lines) and CO(2--1) (dotted lines) spectra of 3C\,403. The data were smoothed to 
channel widths of 44 and 50\,km\,s$^{-1}$, respectively. The zero velocity corresponds to the recessional velocity of the galaxy 
($V_{\rm sys}$ = 17688 \kms).}
\label{pico10}
\end{figure}

\subsubsection{CO spectra and millimeter wave continuum}\label{CO}

Motivated by the detection of megamaser emission, we searched for the two ground rotational transitions of the most common observable 
molecule, CO. Unlike 22\,GHz H$_2$O, the CO\,(1--0) and (2--1) lines predominantly trace cool, low-density gas. No clear 
emission is seen down to 3$\sigma$ noise limits of 13 and 38\,mJy for the two respective transitions (this refers to the unsmoothed spectra 
with 1 and 4\,MHz channel spacing, see Sect.\,\ref{obssect}). We tentatively detect emission at $\sim$ --300\,km\,s$^{-1}$ 
with respect to the systemic velocity (Sect.\,\ref{vref}). Evans et al. (\cite{evans05}) also reported a non-detection of CO toward 3C\,403 
as part of a survey of IRAS radio galaxies. Curiously, also in their spectrum (their Fig.\,1) a highly tentative feature is present at 
about \hbox{--300}\,km\,s$^{-1}$. While the offset is opposite to that expected from optical emission lines (Sect.\,\ref{vref}) and while 
evidence for such a feature is too weak to seriously question the recessional velocity adopted throughout this paper, a CO spectrum 
with higher sensitivity would be desirable. 

Indicated by the baselines, continuum emission was also detected. At 109\,GHz, the level is approximately 25 $\pm 13$\,mJy (the uncertainty was estimated from the variations of the continuum level in individual scans). No continuum was detected at 218\,GHz.  

\subsection{Continuum emission}

\subsubsection{22\,GHz VLA emission}\label{continuosect}

\begin{figure}
\centering
\includegraphics[width= 8 cm]{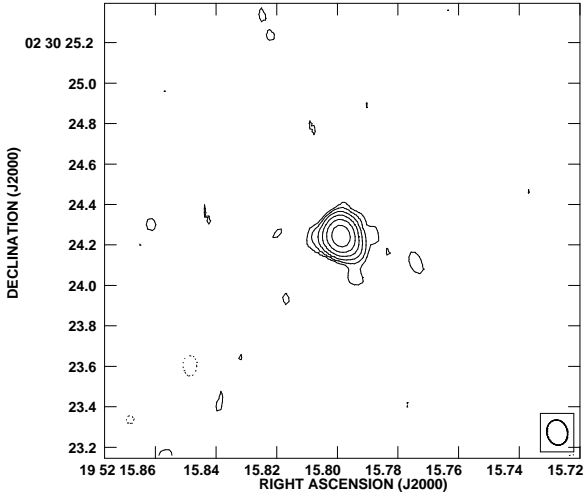}
\caption{A naturally weighted 22\,GHz VLA radio continuum image (resolution: 0\,\farcs1, corresponding to $\sim$115\,pc) of the nuclear 
region of 3C\,403, made using channels free of line emission. The peak is at 40\,mJy\,beam$^{-1}$. The contours are (--1, 1, 2, 4, ..., 32) 
$\times$ 0.75\,mJy\,beam$^{-1}$.}
\label{contvlan}
\end{figure}

\begin{figure}
\centering
\includegraphics[width= 8 cm]{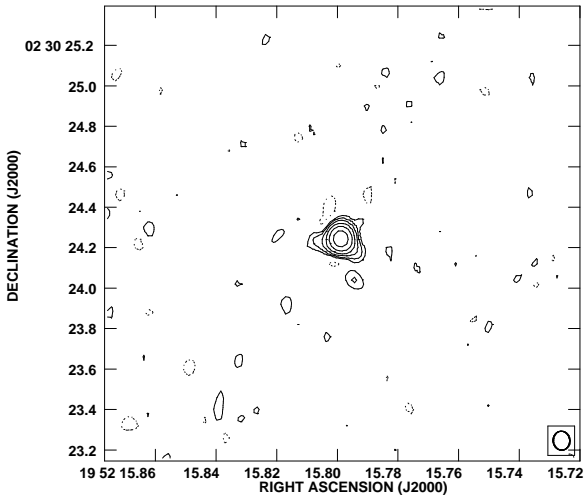}
\caption{A uniformly weighted 22\,GHz VLA radio continuum image (resolution: 0\,\farcs08, corresponding to $\sim$90\,pc) of the nuclear 
region of 3C\,403, made using channels free of line emission. The peak is at 39\,mJy\,beam$^{-1}$. Contours are (--1, 1, 2, 4, ..., 32) 
$\times$ 0.75\,mJy\,beam$^{-1}$.}
\label{contvlau}
\end{figure}

In Fig.\,\ref{contvlan}, we present a naturally weighted VLA A-array 22\,GHz radio continuum image of the nuclear region of 3C\,403, 
produced using maser emission-free channels. The source is spatially unresolved with a peak flux density of 40\,mJy beam$^{-1}$. However, 
an elongation is apparent in the uniformly weighted image of Fig.\,\ref{contvlau} that might be a weak signature of a parsec scale jet 
visible in the higher resolution images presented in Sect.\,\ref{evnsect}. 

\begin{figure}
\centering
\includegraphics[width= 8 cm]{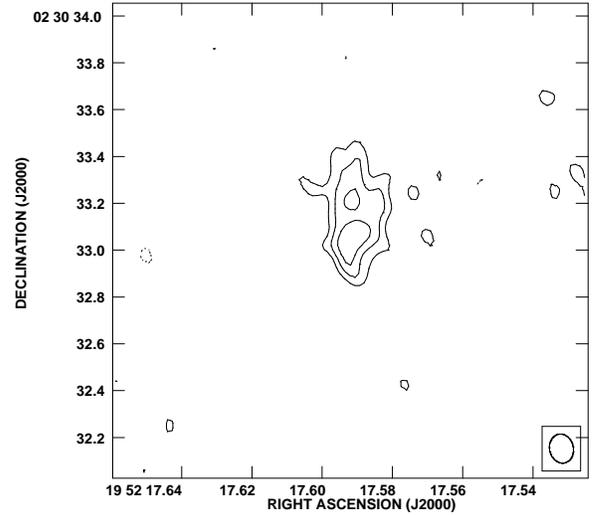}
\caption{A naturally weighted 22\,GHz VLA radio continuum image (resolution: 0\,\farcs1, corresponding to $\sim$115\,pc) of the bright radio knot 
(F6, following DSL) in 3C\,403, obtained from channels free of line emission. The peak is at 1.5\,mJy\,beam$^{-1}$. Contours are 
(--1, 1, 1.5, 2) $\times$ 0.6\,mJy\,beam$^{-1}$.}
\label{contvlak}
\end{figure}

\begin{figure}
\centering
\includegraphics[width= 8 cm]{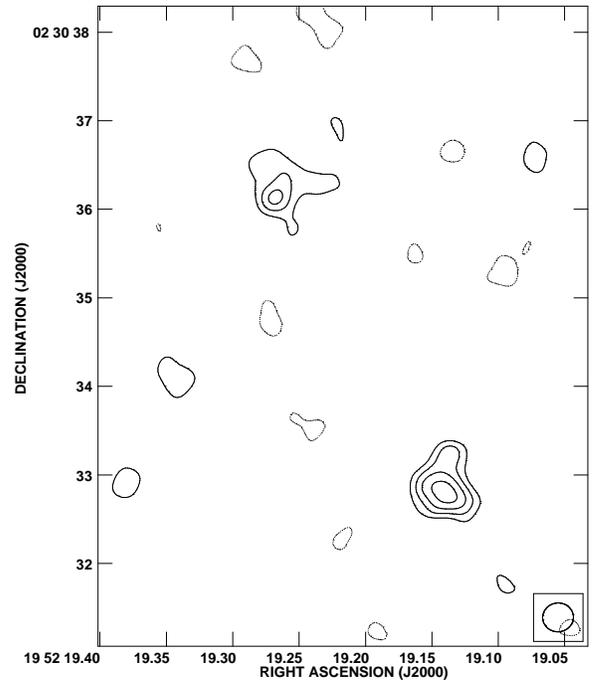}
\caption{A 500 k$\lambda$ tapered 22\,GHz VLA radio continuum image (resolution: 0\,\farcs33, corresponding to $\sim$375\,pc) of the compact 
hotspot structures (F1 and F2, following BBL) in 3C\,403, made using channels free of line emission. The peak is at 
2.5\,mJy\,beam$^{-1}$. The contours are (--1, 1, 1.5, 2, 2.5) $\times$ 0.9\,mJy\,beam$^{-1}$.}
\label{contvlah}%
\end{figure}

As for the H$_2$O line emission, we have also investigated a $\sim$ 3$^{\prime}$ $\times$ 3$^{\prime}$ area in a search for additional 
radio continuum emission. Apart from the nucleus, among the radio continuum features (lobes, radio jets, and bright knots) reported by 
BBL and DSL, we have only detected emission at position RA${\rm _{J2000}}$ = 19$^{\rm h}$ 52$^{\rm m}$ 17.6$^{\rm s}$, 
Dec${\rm _{J2000}}$ = 02$^{\rm \circ}$ 30$^{\rm \prime}$ 33$^{\rm \prime\prime}$, coincident within the errors with the exceptionally 
bright radio knot F6 (nomenclature according to BBL). The knot in our map (Fig.\,\ref{contvlak}) seems to be resolved into two components 
aligned almost perpendicular to the large scale radio jet. Using a 500\,k$\lambda$ taper function, weak emission is also detected at the 
position where two compact structures in the eastern hotspot (F1 and F2, following BBL) have been previously observed 
(Fig.\,\ref{contvlah}). The remaining radio emission from 3C\,403 detected by DSL is either resolved out or is below the detection 
threshold given in Sect.\,\ref{obssect}.

\subsubsection{VLBI maps}\label{evnsect}

\begin{figure}
\resizebox{\hsize}{!}{\includegraphics[angle = 0]{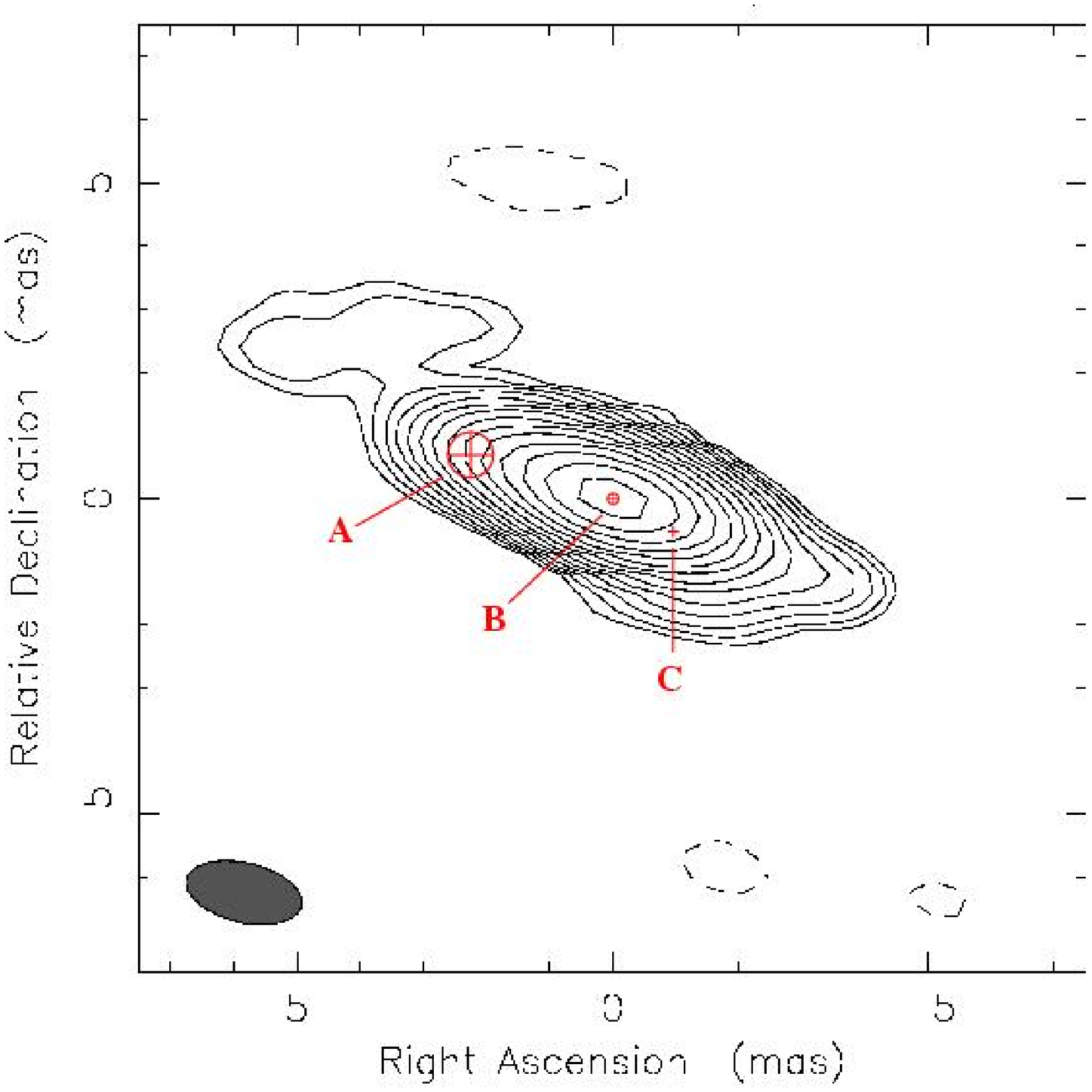}}
\caption{Full resolution EVN image of the nucleus of 3C\,403 at $\lambda$=6\,cm. The contours start at 0.2 mJy and increase with a
factor of 1.5. The beam size of the observations is 1.87\,mas$\times$0.94\,mas at a position angle of 76$\,.\!\!^{\circ}$9.}
\label{original}
\end{figure}

\begin{figure}
\resizebox{\hsize}{!}{\includegraphics[bbllx=2.4cm,bburx=16cm,bblly=7cm,bbury=21cm,clip=, angle=-90]{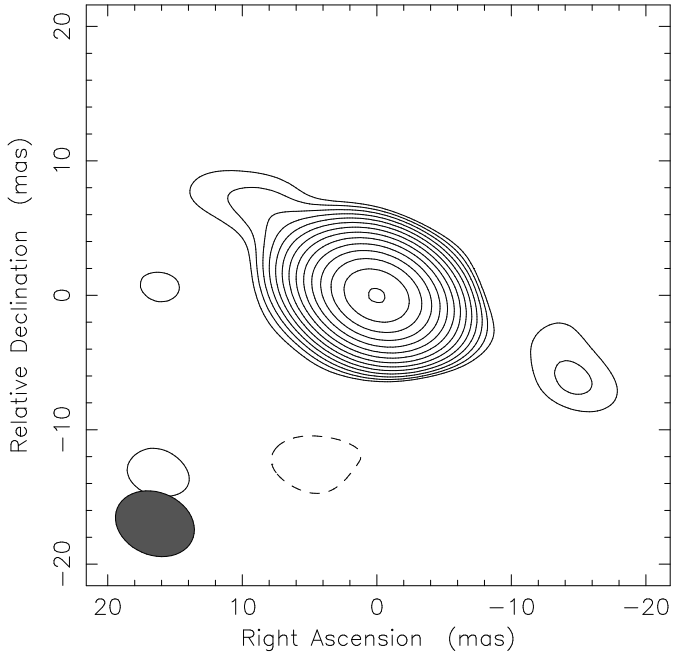}}
\caption{Low resolution EVN image of the nucleus of 3C\,403 at $\lambda$=6\,cm using only the continental EVN antennas. The contours
start at 0.35 mJy and increase with a factor of 1.5. The beam size is 6.05\,mas$\times$4.67\,mas at a position angle of 
67$\,.\!\!^{\circ}$4.} 
\label{inner}
\end{figure}

\begin{figure}
\resizebox{\hsize}{!}{\includegraphics[angle=-90]{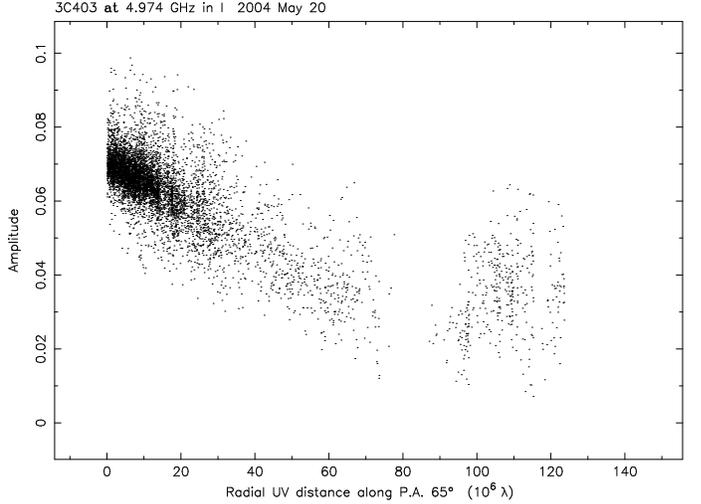}}
\caption{The visibilities of 3C\,403 clearly show that the source is resolved on the longer baselines.}
\label{projplot}
\end{figure}

Milliarcsecond resolution 6-cm VLBI maps of the radio nucleus of 3C\,403 are shown in Figs.\,\ref{original} and \ref{inner}.  
The visibilities of 3C\,403 demonstrate that the source is resolved at longer baselines (Fig.\,\ref{projplot}). The amplitude drops 
from 70\,mJy at the shortest baselines (10\,M$\lambda$) to 40\,mJy at 60\,M$\lambda$.

The uniformly weighted VLBI map of 3C\,403 is shown in Fig.\,\ref{original}. Notable are extensions to the north-east and south-west. 
We fitted circular Gaussian components to the uv-data, and the best fit with three components gave a reduced $\chi^2$ per degree
of freedom of $\sim$ 1.7. Reducing the number of components to two or one increases the reduced $\chi^2$ p.d.f to 2.2 and 2.9,
respectively. On the other hand, adding more components does not significantly improve the fit. Hence, we used a three component model 
and the results from this fit are shown in Table~\ref{model_tab}. We  fitted the three components many times with different 
initial parameters and used the scatter in the results to estimate the uncertainties of the fit parameters. While the flux densities 
and angular sizes of the three components are not strongly constrained by the fit, the position angle is well determined. The 
orientation of the pc-scale jet is in very good agreement with the kpc-scale jet seen on VLA images.

Since the telescope in Urumqi (China) could not take part in the observations, all information on the long baselines is based on 
only two telescopes, Hartebeesthoek (South Africa) and Shanghai (China). Hence the data on long baselines, mainly responsible for 
the uncertainties in the model fit parameters in Table \ref{model_tab}, should be considered with some caution. To test the result
shown in Fig.\,\ref{original} we also self-calibrated and imaged the data from the short baselines only. As before, the amplitude 
in the visibilities shows a slow decrease towards longer baselines. The resulting map is shown in Fig.\,\ref{inner}. The extension 
toward the north-east is visible in this map and there might be also a weak jet component towards the south-west. From this we 
conclude that the weak extensions on mas scales are real and are not caused by amplitude calibration errors.

\begin{table}
\caption[]{Model fit of the three components A, B, and C to the 
uv-data. $S$ is the flux density, $d$ denotes the separation from 
the phase center.}
 \label{model_tab}
\[
\begin{tabular}{p{0.14\linewidth}|p{0.16\linewidth}p{0.16\linewidth}p{0.16\linewidth}cp{0.16\linewidth}}
   \hline
Comp. & $S$ [mJy] & $d$ [mas] & Size [mas]& P.A. \\
    \hline
	A		&  9.4$\pm$5& 2.41$\pm$0.6& 0.69$\pm$0.20 &  72.5$\pm$3  \\
	B		& 45.2$\pm$10& 0.00        & 0.15$\pm$0.05 &   0.0  \\
	C		& 14.7$\pm$7& 1.02$\pm$0.3& $<$ 0.04 &--117.1$\pm$5  \\

    \hline
 \end{tabular}
\]
\end{table}

Assuming the presence of a core and one or two nuclear jet components, there are three interpretations: 
\begin{itemize}
\item Component A is the nucleus. Then we are dealing with a one sided jet towards the south-west.
\item Component B is the nucleus and components A and C are jet and counterjet.
\item Component C is the nucleus and a one sided jet goes towards the north-east.
\end{itemize}

\section{Discussion}

\subsection{Evidence for an edge-on torus}\label{torus}

\subsubsection{The radio continuum}

To select the most likely of the three scenarios outlined in Sect.\,\ref{evnsect} and to identify the location of the radio nucleus, 
we compare our continuum VLBI maps with the large scale structure of 3C\,403. The orientation of component A (P.A.=73$^\circ$; 
Table~\ref{model_tab}) is in very good agreement with the eastern large scale jet and the position of a bright knot (73$^\circ$) in 
its lobe (e.g., BBL, DSL). The south-western jet (component C), as seen in Fig.\,\ref{original}, seems to bend slightly to the west, 
thus pointing towards the western large scale jet. In view of the limited uv-coverage w.r.t. long baselines (see Sect.\,\ref{evnsect}), 
we consider this latter coincidence as tentative. 

The presence of the western and eastern large scale jets with similar flux densities suggests that the jets must lie close to the plane 
of the sky. The inner 10\,pc of 3C\,403 (Figs.\,\ref{original} and \ref{inner}) show a similar morphology. Thus a scenario with B 
representing the core and A and C marking jet and counterjet is by far the most likely one. If we see the jet and the counterjet with similar strengths, the orientation of the jets must be close to the plane of the sky. This view is supported by Kraft et al. 
(\cite{kraft05}; hereafter K05). They present results from {\it Chandra} X-ray observations and describe the X-ray spectrum of 
the nucleus as a superposition of two power-law continuum components with a 6.4\,keV Fe line. One of these power-law components is 
heavily absorbing ($N_{\rm H}$ $\sim$ 4$\times$10$^{23}$cm$^{-2}$). 

If the X-ray photons are predominantly scattered radiation not passing through the main body of the obscuring torus, the high column 
density derived by K05 would only be a lower limit. Since no evidence was found for a significant amount of dust in observations with 
the Hubble Space Telescope (Martel et al. \cite{martel99}), K05 conclude that the absorption is due to material close to the nuclear 
engine, possibly forming a molecular torus. Spectroscopic studies at optical wavelengths revealed narrow emission lines (Tadhunter et 
al. \cite{tadhunter93}), indicating that the broad line region is hidden behind an edge-on dusty torus.

\subsubsection{On the absence of H{\sc i} and CO}

Morganti et al. (\cite{morganti01}) searched for H{\sc i} absorption in 3C\,403 and got an upper limit of $N_{\rm H}$ $<$ 
5$\times$10$^{20}$\,cm$^{-2}$ ($T_{\rm spin}$ = 100\,K). Large differences in H{\sc i} and X-ray column densities are quite common in Seyfert galaxies (e.g. 
Gallimore et al. \cite{gallimore99}). The striking discrepancy between X-ray (Sect.\,5.1.1) and H{\sc i} column densities indicates 
the predominance of either ionized or molecular gas along the line of sight toward the nuclear source of 3C\,403. While both kinds 
of environment may coexist in a small volume surrounding the central engine, the maser emission provides substantial evidence for 
the latter. 

Why are we then not seeing CO? For a quasi-thermally excited line, seen in emission, beam dilution may become important in an object
as distant as 3C\,403. With the observed brightness temperature as a function of the intrinsic one as well as source and beam solid 
angle ($T_{\rm b}^{\rm obs}$ = [$T_{\rm b}^{\rm int}$ $\times$ $\Omega_{\rm torus}$]/[$\Omega_{\rm torus}$ + $\Omega_{\rm beam}$]) 
we can determine the smallest toroid size, $\Omega_{\rm torus}$, we would have detected. With a CO line detection threshold of 
$T_{\rm mb}^{\rm obs}$ = 1.8\,mK and a 109\,GHz beam size of 23\arcsec\ (Sect.\,2), we obtain $\Omega_{\rm torus}$ $\sim$ 0\ffas04 
$\times$ (500\,K/$T_{\rm b}^{\rm int}$)$^{1/2}$ that corresponds to a linear scale of $\sim$ 50\,pc $\times$ 
(500\,K/$T_{\rm b}^{\rm int}$)$^{1/2}$. Elliptical galaxies are characterized by high stellar velocity dispersions efficiently
heating any cool gas and creating an X-ray emitting plasma that is greatly confining the volume occupied by molecular clouds
(e.g., Wiklind \& Henkel \cite{wiklind}). The lack of dust emission (Martel et al. \cite{martel99}) supports a small size of the 
molecular complex in the nuclear region and is therefore consistent with our non-detection of CO.

For the absence of deep CO $J$=1--0 absorption against the nuclear radio continuum source (the absorption must reach a significant
fraction of the continuum level, otherwise it would not be detected; see Sect.\,\ref{CO} and Fig.\,\ref{pico10}), there are 
several possible explanations. (1) There is no molecular gas in front of the nuclear source. In view of the column density 
(K05) this is not likely. (2) The torus is partly or entirely molecular but has a very small fractional CO column density 
in the $J$ = 0 and 1 states. This may be a consequence of a high temperature of the gas, possibly coupled with an intense radiation 
field (resulting in excitation temperatures of several 100 to $\sim$ 2000\,K). (3) The obscuring clouds may be 
much smaller than the radio continuum source, similar to what was proposed for the much stronger radio emitter Cyg~A by Barvainis 
\& Antonucci (\cite{barvainis94}). Within this context the size of the H$_2$O maser hotspots (Sect.\,4.1.1) does not provide any 
guideline, because 22\,GHz H$_2$O emission may originate from much higher density gas then the ground rotational lines of CO. (4) Alternatively, it is also possible that any existing absorption would be `swamped' by emission at the same velocity interval.

\subsection{Origin of the H$_2$O maser emission}

\subsubsection{Constraints on the application of the standard model}

The main result of our VLA observations is that the two 22\,GHz H$_2$O velocity components of 3C\,403 do not arise from the extended 
lobes of the system, but are aligned with the central radio continuum source within a $\sim$5.5 pc sized region. Furthermore, the 
nuclear torus likely has an edge-on orientation (see Sect.\,\ref{torus}). Thus an association of the maser features with a 
dense circumnuclear torus/accretion disk is a possibility worth discussing. 

To analyze this case, we have to evaluate the statistical properties of the sample of known megamaser galaxies. Accounting for the 
sources given in Henkel et al. (\cite{henkel05}), Kondratko et al. (\cite{kondratko06}), and Zhang et al. (\cite{zhang06}), the latter 
containing 64 galaxies with detected H$_2$O emission beyond the Magellanic Clouds, there are a total of 51 known megamaser sources 
($L_{\rm H_2O}$ $>$ 10\,L$_{\rm \odot}$). Most of these have not yet been studied in spatial detail. Omitting 3C\,403, the sample 
contains two pure jet-masers (apparent interaction between a nuclear jet and a dense molecular cloud), while 16 sources appear to 
be disk-masers arising from an accretion disk. The latter is either based on detailed observations as in the case of NGC\,4258 
(e.g., Miyoshi et al. \cite{miyoshi95}) or, a little more speculative, on the measured lineshape (see, e.g., NGC\,6323 in Braatz et 
al. \cite{braatz04}). H$_2$O megamaser emission associated with a nuclear outflow is only known for one source of the sample, the 
Circinus galaxy. Even here, however, only a part of the emission is associated with the outflow, complementing a nuclear accretion 
disk viewed approximately edge-on (Greenhill et al. \cite{greenhill03}).    

In view of the known sample of megamaser galaxies, an association of the maser in 3C\,403 with a nuclear outflow appears to 
be unlikely. So far all detected `jet-maser' components are observed at one side w.r.t. the systemic velocity (Henkel et al. 
\cite{henkel05}). We should keep in mind,however, that 3C\,403 is an (elliptical) FR~II and not a (spiral) Seyfert 2 or LINER 
galaxy and that "disk-masers" may be identified more easily than "jet-masers". Nevertheless, observed lineshapes and the relatively large number of disk masers superficially hint toward H$_2$O emission from an accretion disk as the most plausible interpretation of the H$_2$O lines from 3C\,403.  

If the emission were part of an accretion disk, at which galactocentric radius would it arise? Assuming that the two spectral 
components originate from the tangentially seen parts of a nuclear accretion disk seen approximately edge-on ($i$ $\sim$ 90$\degr$), 
we obtain (Fig.\,\ref{monitor}) a rotation velocity of $V_{\rm rot} = V_{\rm obs}\,\sin^{-1} i$ $\sim$100\,km\,s$^{-1}$. From 
Bettoni et al.\ (\cite{bettoni03}; their Table 3) we obtain $M_{\rm BH}$ = 10$^{8.8}$\,M$_{\odot}$ as the mass  of the nuclear engine. 
Combining these two parameters and assuming the  presence of Keplerian rotation (as in NGC\,4258, see Miyoshi et al. \cite{miyoshi95};
Herrnstein et al. \cite{herrnstein99}, \cite{herrnstein05}), this yields with  
$$
R = 0.89 \left[\frac{M_{\rm BH}}{\rm M_{\odot}}\right]
 \left[\frac{V_{\rm obs}\,\sin^{-1} i}{\rm km\,s^{-1}}\right]^{-2} 
 \left[\frac{D}{\rm Mpc}\right]^{-1} {\rm mas}
$$
an angular distance of 245\,mas, corresponding to a galactocentric radius of $R_{\rm GC}$ $\sim$ 275\,pc. This size conflicts with the result of our VLA observations (Sect.\,\ref{vlares}) that, indicating that the continuum and the H$_2$O emission are arising from the same 5.5\,pc (5\,mas) sized region, imply a maximal galactocentric radius for the H$_2$O masers of 2.75\,pc (2.5\,mas). 

Is it possible to reconcile this value with the possibility that the maser arises in an accretion disk like that in NGC\,4258?  A 2.75-pc radius disk rotating at 100\,km\,s$^{-1}$ would imply a black hole mass $M_{\rm BH}$ = 10$^{6.8}$\,M$_{\odot}$, a factor of 100 lower than that derived by Bettoni et al. \cite{bettoni03}. Alternatively, we might consider a disk inclined by $\sim$6$\degr$ or less. However, this would describe an almost face-on disk, a scenario not only different from that of NGC\,4258 but also seemingly contradicting the presence of jets oriented approximately along the plane of the sky. 

To summarize, {\it the masers in 3C\,403 are peculiar in three different ways. (1) They are detected at the core of a radio galaxy.
(2) The two main ultraluminous velocity components are highly time variable and (3) the Keplerian model seems to fail}. If the main 
velocity components do not represent the two tangentially viewed parts of the masing disk but instead only one part and the systemic 
features, this would lead to $V_{\rm rot}$$\sim$200\,km\,s$^{-1}$ and with $M_{\rm BH}$ = 10$^{8.8}$\,M$_{\odot}$ (Bettoni et al.\ \cite{bettoni03}) to a galactocentric distance of 60\,mas (70\,pc). In this case 
the disagreement with observations is slightly less severe but the linear scale is still inconsistent with the results from our VLA 
A-array measurements (Sect.\,\ref{vlares}). Thus, the accretion disk scenario requires relatively strict constraints to be applied to our case unless we are only viewing the systemic velocity features of the putative maser disk (see below).

\subsubsection{Alternatives to the accretion disk scenario}

{\bf {Cloud stability considerations}}: A main difference between Seyfert and FR~II galaxies (equivalent to radio loud type 2 
quasars according to the standard unified scheme) is the mass of the nuclear engine. While in Seyfert galaxies masses range from 
$\sim$10$^6$ to a few 10$^7$\,M$_{\odot}$ (e.g., Greenhill et al. \cite{greenhill96}, \cite{greenhill97}; Herrnstein et al. 
\cite{herrnstein99}, Henkel et al. \cite{henkel02}), masses in radio galaxies and quasars reach 10$^9$\,M$_{\odot}$, with 
3C\,403 hosting a nucleus that is not far below this latter value (Bettoni et al. \cite{bettoni03}). Could clouds in an NGC\,4258-like 
disk that is characterized by a particularly large rotational velocity and small galactocentric radius be stable in an environment 
dominated by such a large central mass?

The masers in NGC\,4258 might arise from well confined discrete clumps that are encompassing only a small fraction of the entire 
volume of the warped circumnuclear disk. For a roughly spherical clump at distance $R_{\rm GC}$ from the central engine and
neglecting magnetic fields (see Modjaz et al. \cite{modjaz05}), a density of 
$$
n({\rm H_2}) \ga\ \frac{3}{2\,\pi\,{\rm G}} \ \left( \frac{V_{\rm rot}}{R_{\rm GC}} \right)^2
$$
is required to reach stability against tidal disruption (Stark et al. \cite{stark89}). G is the gravitational constant. For 
NGC\,4258, i.e. $R_{\rm GC}$ $\sim$ 0.2\,pc and $V_{\rm rot}$ $\sim$ 1000\,km\,s$^{-1}$ (Herrnstein et al. \cite{herrnstein99}, 
\cite{herrnstein05}), the resulting density becomes $n$(H$_2$) $\sim$ 5$\times$10$^{10}$\,cm$^{-3}$. This is close to thermalization 
(Kylafis et al. \cite{kylafis91}). For commonly accepted collisional excitation this either hints at excitation by a two temperature 
gas (e.g., Kylafis et al. \cite{kylafis87}) or at emission from turbulent, warm, dense molecular debris that does not form self-gravitating
clouds. 

Most circumnuclear maser disks are characterized by smaller rotation velocities and higher galactocentric distances, significantly 
reducing the required density for stability against tidal disruption so that here the assumption of long lived masing clumps giving
rise to collisionally excited masers is less of a problem (e.g., $n$(H$_2$) $\ga$ 10$^8$\,cm$^{-3}$ in NGC\,1068; see Greenhill et al. 
\cite{greenhill96}; Gallimore et al. \cite{gallimore01}). Assuming for a putative accretion disk in 3C\,403, where only the systemic 
spectral features are seen (Sect.\,5.2.1), $R_{\rm GC}$ = 0.2\,pc (as in NGC\,4258), the corresponding minimum density for cloud 
stability reaches, however, 10$^{12}$\,cm$^{-3}$, which is prohibitively large for collisionally excitated maser emission by a one 
temperature gas. Could then a lack of stable, well ordered clouds explain the rapid variability of the maser features in 3C\,403? At 
the maximal galactocentric distance permitted by our VLA data, $R_{\rm GC}$ $\sim$ 2.75\,pc (Sect.\,4.1.2), the required density would drop to more reasonable $\ga$10$^{8}$\,cm$^{-3}$. 

\hfill\break\noindent
{\bf {The jet-maser scenario}}: While in Seyfert and LINER galaxies statistics favor accretion disk over jet-masers (Sect. 5.1.1), the 
two main H$_2$O velocity components are (1) quite broad and (2) offset by several 10\,km\,s$^{-1}$ from the nominal (NED) systemic 
velocity of the galaxy. These are the typical spectral properties of jet masers (Peck et al. \cite{peck03}; Henkel et al. \cite{henkel05}). 
The flares of velocity components I and II (Fig.\,\ref{spectrum}) may arise from shocked regions at the interface between the energetic
jet material and the molecular gas of the cloud the jet is boring through. The time delay between components I and II (Fig.\,\ref{monitor})
and the apparent red- (component I) and blue-shift (component II) with respect to the systemic velocity may be caused by two clouds at 
opposite sides of the nucleus that have a slightly different galactocentric radius, possibly being part of a circumnuclear molecular ring.

\hfill\break\noindent
{\bf A binary supermassive black hole in 3C403?} Aside from the mass of the nuclear engine, there may be another fundamental difference between 
3C\,403 and NGC\,4258 that is worth mentioning. On a large spatial scale, 3C\,403 shows not the standard two-sided radio jet, but a 
pair of such jets (see Sect.\,1).  The apparently younger one is oriented roughly NE-SW, the more extended, weaker one in the SE-NW direction. 
It is thus one of the rare examples of an X-shaped radio galaxy (e.g., Figs.\,1 and 2 of Kraft et al. \cite{kraft05}). This morphology is 
believed to be caused by the coalescence of two supermassive black holes (e.g., Gopal-Krishna et al. \cite{gopal}; Liu \cite{liu}) whose 
interaction modifies the inclination of the pre-existing nuclear disk of the dominant galaxy. Such a scenario may lead to a complexity that 
is much higher than that encountered in the nuclear environment of NGC\,4258, yielding a puzzling number of possible disk- and jet-maser 
configurations that have to be constrained by future measurements. In view of the number of free parameters in such an environment, it 
remains to be seen whether the anticorrelation of the main two maser features was a rare accidental event or whether it will provide an 
important clue for a better understanding of the so far poorly explored circumnuclear regions of X-shaped radio galaxies.

\section{Conclusions and outlook}

We have investigated the luminous megamaser in the X-shaped FR~II galaxy 3C\,403 using single-dish monitoring of the line 
features and VLA A-array observations. The two main maser features show extreme intensity variations during a two-year period. The 
position of the maser as derived from the VLA maps is consistent with a location within a few pc of the radio nucleus of the galaxy. 

In addition, we have used the EVN to study the galaxy's nuclear radio continuum on parsec scales. A dominant central component and two 
extensions toward the south-west and north-east are identified. Because these parsec-scale extensions show position angles that are 
compatible with those of the large scale jets and because these large scale jets show similar flux densities, an interpretation in 
terms of a dominant nuclear core and a two-sided jet is plausible. 

No clear sign of CO emission or absorption has been detected. If emission and absorption are not accidentally canceling each other, this
limits the extent of a low density molecular gas component to a size of order $\la$200\,pc.

If the mass of the nuclear engine is as large as previously estimated and if the nuclear accretion disk is oriented (as expected) almost 
edge-on, an NGC\,4258-like accretion disk scenario is only viable, if all the detected H$_2$O features are systemic, arising from the front 
or back side of the disk. There is, however, no observational evidence for the tangentially viewed approaching and receding parts of the 
disk. A secular velocity drift of the observed components is also not seen. In view of linewidths and deviations from the systemic 
velocity, an interpretation in terms of "jet-masers", arising from molecular clouds shocked by the nuclear jet(s), is the most plausible 
one. 

Since 3C\,403 is an X-shaped radio galaxy, it may host a binary nuclear engine, greatly enlarging the number of potential scenarios 
that should be observationally constrained by future measurements. These imply single-dish maser monitoring over long time spans (1) to 
study the lifetime of the various components, (2) to check whether the anticorrelation of the two main features is a typical or an accidental
event and (3) to find, possibly, some evidence for a systematic velocity drift. The latter would be a clear sign for the presence of a nuclear 
disk even if the red- and blue-shifted parts of this disk remain undetectable (see Wilson et al. \cite{wilson95} and Braatz et al. \cite{braatz03} for such a previously identified case). Another very important measurement would be the simultaneous observation of the maser features and the radio continuum at \hbox{(sub-)}milliarcsecond resolution. This could directly show whether 
the maser features are associated with the nuclear jets or whether we should reject this scenario. In any case it would shed light onto 
the morphology of the nuclear environment of a prototypical X-shaped galaxy. Finally, systematic monitoring of the radio continuum, 
H$_2$O and X-ray spectra may provide through reverberation mapping important correlations that would be essential to assess the physical 
state and three-dimensional morphology of the circumnuclear medium surrounding a 10$^{8.8}$\,M$_{\odot}$ AGN.

\begin{acknowledgements}
AT would like to thank Karl Heinz Wissmann and his collaborators for pleasant discussions during the making of this work. AB is supported by the Priority Programme 1177 of the Deutsche Forschungsgemeinschaft. We wish to thank J. Moran for critical comments. This research has made use of the NASA/IPAC Extragalactic Database (NED) which is operated by the Jet Propulsion Laboratory, California Institute of Technology, under contract with the National Aeronautics and Space Administration. We have also made use of NASA's Astrophysics Data System.    
\end{acknowledgements}


\section*{APPENDIX: 4C\,+02.49}
\label{4c249}

In order to find suitable phase calibrators for VLBI observations, we observed nine bright and compact sources from 
the NRAO VLA Sky Survey (NVSS) with an angular separation of $<$2$^{\circ}$ with respect to 3C\,403. The 22\,GHz 
($\lambda$$\sim$1.3\,cm) observations were carried out with the VLA on February 5, 2004. The total observing time was 
1.5 hours yielding $\sim$4 minutes on-source integration time per source. The data reduction was performed using standard 
procedures in the Astronomical Image Processing System (AIPS). The most suitable source w.r.t. morphology and flux 
density turned out to be 4C+02.49 (see Sect.\,2).

\begin{figure}
\resizebox{\hsize}{!}{\includegraphics[angle=0]{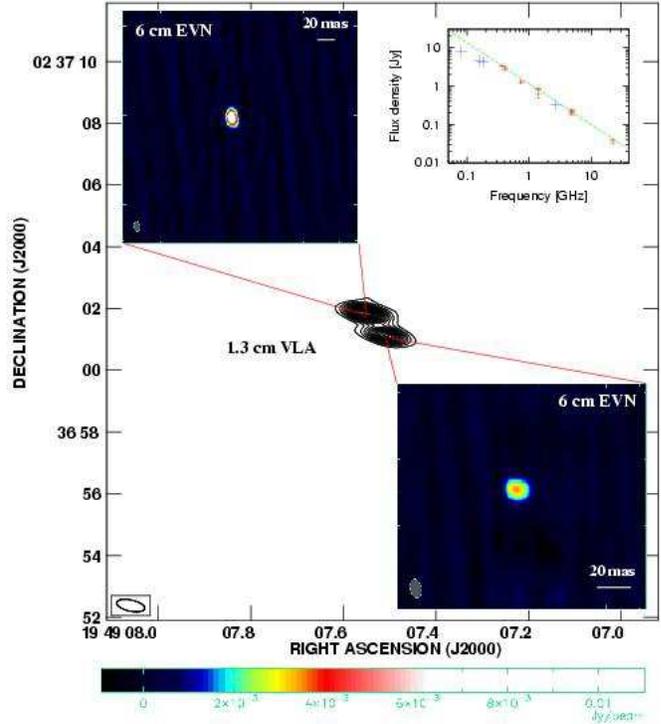}}
\caption{VLA map at 1.3 cm (center) and EVN images (upper left and lower right) of the two components of 4C\,+02.49. 
Also shown is a broad band radio spectrum (upper right) obtained from NED and our VLA 22\,GHz flux. The large (blue) crosses 
indicate flux density measurements that had no error bar.}
\label{1949}
\end{figure}

The VLA 1.3\,cm observations show that 4C+02.49 is a compact double with flux densities of 21.7$\pm$0.5 and 15.3$\pm$0.5\,mJy 
for the north-eastern and south-western component, respectively (Fig.\,\ref{1949}). A broad band radio spectrum of 
4C+02.49 (see the inset in Fig.\,\ref{1949}) with total integrated flux densities obtained from the NASA Extragalactic Database 
(NED) and our VLA observation gives a steep spectrum with spectral index of --1.07$\pm$0.03.  Hence, this source should be 
classified as a compact steep spectrum (CSS) source (Fanti et al. \cite{fanti01}). The tight correlation between all data 
points indicates that there is little variability in the total flux. According to the turnover-frequency vs. linear size 
relation in CSS sources (O'Dea \& Baum \cite{odea97}), one expects a source size of $>$10\,kpc for a turnover frequency 
$<$100\,MHz. This cannot be tested, since, so far, no redshift has been derived for this source.  

The two compact components of 4C+02.49 were clearly detected with the EVN at $\lambda$$\sim$6\,cm (Fig.\,\ref{1949}). The 
flux densities are 18.4$\pm$0.6 (upper left) and 8.8$\pm$0.6\,mJy (lower right). Hence, most of the emission that is 
responsible for the total integrated (NED) 6\,cm flux of $\sim$220\,mJy is resolved out. This is likely the main cause 
for the seemingly inverted spectra when comparing the EVN 6\,cm with the VLA 1.3\,cm flux densities.
 
\end{document}